\documentclass[pdflatex,sn-mathphys-num]{sn-jnl}

\usepackage{graphicx}%
\usepackage{multirow}%
\usepackage{amsmath,amssymb,amsfonts}%
\usepackage{amsthm}%
\usepackage{mathrsfs}%
\usepackage[title]{appendix}%
\usepackage{xcolor}%
\usepackage{textcomp}%
\usepackage{manyfoot}%
\usepackage{booktabs}%
\usepackage{algorithm}%
\usepackage{algorithmicx}%
\usepackage{algpseudocode}%
\usepackage{listings}%
\usepackage{subfigure}

\theoremstyle{thmstyleone}%
\theoremstyle{thmstyletwo}%

\theoremstyle{thmstylethree}%

\begin{document}

\title[Article Title]{High-power beyond extreme ultraviolet FEL radiation with flexible polarization at SHINE}


\author{\fnm{Hanxiang} \sur{Yang}}
\author{\fnm{Zhangfeng} \sur{Gao}}
\author{\fnm{Bingyang} \sur{Yan}}
\author{\fnm{Wencai} \sur{Cheng}}
\author{\fnm{Nanshun} \sur{Huang}}
\author*{\fnm{Haixiao} \sur{Deng}$^{*}$}

\affil{\orgdiv{Shanghai Advanced Research Institute}, \orgname{Chinese Academy of Sciences}, \postcode{{201210}, \state{Shanghai}, \country{China}}}

\email{denghx@sari.ac.cn}


\abstract{Linac-based free-electron lasers (FELs) feature high brightness, narrow bandwidth, controllable polarization, and wide wavelength tunability. With the rapid development of superconducting radio-frequency technology, linacs can now operate at MHz-level repetition rates, enabling FELs with both high repetition rates and high average power. Beyond extreme ultraviolet (BEUV) radiation is of great interest for scientific research and industrial applications, especially for next-generation lithography. Owing to the main design parameters of SHINE, the generation of BEUV radiation is a natural capability of the facility. The BEUV characteristics at SHINE are investigated and its achievable performance as a high-average-power light source is evaluated. By applying undulator tapering to enhance the energy extraction efficiency, kilowatt-level BEUV radiation with controllable polarization is shown to be achievable. These results demonstrate that SHINE can provide a high-performance BEUV source, offering a realistic pathway toward a high-average-power light source for next-generation high-resolution lithography.}

\keywords{free-electron laser, beyond extreme ultraviolet, lithography}



\maketitle

\section{Introduction}\label{sec1}
Beyond extreme ultraviolet (BEUV) light sources, operating at wavelengths around 6.X nm, are regarded as a key enabling technology for next-generation nanolithography and advanced photon-based scientific research \cite{Fu2019}. According to the Rayleigh criterion \cite{harriott2002limits}, reducing the exposure wavelength is one of the most effective approaches to improve lithographic resolution \cite{Tallents2010EUV,fomenkov2017light}. As lithography at 13.5 nm approaches its practical limits, extending the operational wavelength toward the BEUV regime has emerged as a promising route for technology nodes below 3 nm \cite{nano11112782,endo2014extendibility}. In addition to wavelength scaling, a sufficiently high average photon flux is essential for high-volume manufacturing, as it directly determines the exposure throughput and helps suppress stochastic defects arising from limited photon dose. In current EUV lithography systems, achieving kilowatt-level average power has been a major technological milestone toward industrial deployment \cite{Fomenkov2019EUV}. Following this trend, a BEUV source with average power at or above the kilowatt level is widely considered a critical requirement for enabling practical next-generation lithography.

Plasma-based light sources, such as laser-produced plasma (LPP) systems \cite{Fujimoto2021,YANG2022100019}, generate unpolarized radiation through unresolved transition arrays from highly ionized high-$Z$ elements. While such sources have been successfully deployed at longer wavelengths, their extension into the BEUV regime faces substantial challenges. The narrow reflectivity bandwidth of multilayer mirrors, which typically requires the BEUV source bandwidth to be below 1\% \cite{Naujok2015,Liu2025tj}, together with the demanding requirements on target materials and debris mitigation, severely limits the achievable BEUV output power and long-term operational stability \cite{nano11112782,Liu2025tj}. These limitations motivate the exploration of accelerator-based light sources, which fundamentally differ from LPP light sources and offer a promising pathway toward kilowatt-level average BEUV radiation \cite{zhao2010,Huang2021,Nakamura2023}.

Storage-ring-based light sources can deliver radiation with broad wavelength coverage by exploiting electron beams operating at repetition rates on the order of 100~MHz \cite{zhao2010}. To further enhance the average power and improve temporal coherence, several advanced concepts have been proposed, including steady-state microbunching \cite{Ratner2010,deng2021experimental,Deng2024}, storage-ring-based self-amplified spontaneous emission (SASE) \cite{Lee2020,hephotonics12100983}, and external laser seeding approaches \cite{Deng2013,Feng2017,Li2020}. Although these schemes show potential for producing high-average-power coherent radiation, extending them into the BEUV spectral range remains highly challenging. In most scenarios, they rely on the manipulation of the electron beam, which can degrade beam quality. Laser-induced energy modulation and radiation processes tend to increase the energy spread, while dispersive sections and complex beam transport can introduce emittance growth \cite{ratner2011reversible,Li2023,Liu2025}. Such degradation limits the FEL gain and extraction efficiency for BEUV generation and makes stable multi-turn operation difficult to maintain.

Linac-based free-electron lasers (FELs), in contrast, provide intrinsically high peak current and superior beam quality, making them well suited for generating high-brightness, fully coherent radiation at short wavelengths \cite{Huang2021}. Among various FEL configurations, SASE FELs \cite{BONIFACIO1984251,SALDIN2002169} are conceptually simple and capable of delivering radiation with excellent transverse coherence, high peak brightness, and flexible polarization control. The average output power of a BEUV source is constrained by the electron beam repetition rate and the single-pulse energy. Recent advances in superconducting radio-frequency (SRF) accelerator technology have substantially expanded the accessible parameter space for high-average-power BEUV FELs. Continuous-wave (CW) SRF linacs can operate at repetition rates reaching the megahertz level, enabling new regimes of FEL operation. Energy-recovery-linac-based FELs (ERLs) offer an efficient route toward high-average-power EUV and BEUV sources by significantly reducing RF power consumption through energy recovery while delivering high-brightness, high-power electron beams \cite{PAGANI2001391,Nakamura2024,zhao2021energy,Kato2019EUVFEL,he2025cavity}. For kilowatt-level BEUV generation at comparable beam energies, the extraction efficiency of ERL-based and single-pass FELs is similar. At higher average power levels (e.g., on the order of 10 kW or above), ERLs are expected to achieve significantly higher wall-plug efficiency, making them attractive for compact and industrial-scale lithography applications. However, ERLs still face challenges in operational complexity, particularly in maintaining high-quality electron beams under high-current conditions. In contrast, single-pass FELs driven by high-quality electron beams at MHz repetition rates provide higher single-pulse energy and enhanced FEL gain efficiency, enabling efficient energy extraction and kilowatt-level BEUV output power. This regime supports a unique combination of high pulse energy and high repetition rate, which is particularly relevant for exploring the performance limits of BEUV FELs.

Representative single-pass FEL facilities include LCLS-II \cite{Stohr2011} and SHINE \cite{Zhao:FEL2017}, both designed for repetition rates approaching 1 MHz and driven by normal-conducting very-high-frequency photoinjectors. Recent progress at SLAC has demonstrated that such linacs can already deliver high-energy, high-brightness electron beams at repetition rates of up to 100 kHz, providing strong experimental validation of this approach \cite{lcls2025rep}. The SHINE facility \cite{Liu2021}, currently under construction and based on similar technologies, can serve as a flexible test platform for exploring BEUV source performance over a wide range of average powers, providing valuable insights into power scaling, source stability, and system optimization for future dedicated industrial light sources. In particular, such a facility offers a complementary research environment to industrial source development, enabling systematic studies toward extreme-resolution lithography.

In this paper, we investigate the feasibility and performance of generating high-average-power BEUV radiation at SHINE based on its main design parameters. Comprehensive start-to-end FEL simulations are carried out to evaluate the baseline BEUV performance and to analyze the effects of energy chirp, undulator tapering, transverse power density distribution, and multi-shot stability. A polarization control scheme employing the elliptically polarized undulator (EPU) afterburner is further proposed to enable flexible generation of linearly and circularly polarized BEUV radiation. By optimizing the undulator tapering to enhance the energy extraction efficiency, kilowatt-level average BEUV output power with millijoule-level pulse energy is shown to be achievable. These results demonstrate that SHINE provides a realistic and promising platform for a high-performance BEUV light source, which is highly attractive for next-generation high-resolution lithography and advanced scientific applications.

\section{BEUV setup at SHINE}\label{sec2}
A BEUV setup based on the layout and main parameters of the Shanghai High Repetition Rate XFEL and Extreme Light Facility (SHINE) \cite{Wang2025,chen2025ultra} is presented. It exploits the high beam quality and repetition rate enabled by SRF linac technology to achieve high-average-power BEUV radiation. The overall working principle and key beam and undulator parameters are introduced to demonstrate the feasibility of the scheme.

\begin{figure}[!htbp]
\centering
\includegraphics[width=1\textwidth]{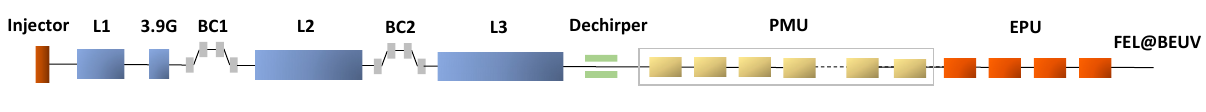}
\caption{Schematic layout of the BEUV setup at SHINE. L1, L2, and L3 denote the linac sections; BC1 and BC2 represent the bunch compressors; 3.9G refers to the 3.9 GHz third-harmonic RF cryomodule section; PMU denotes the planar undulator; and EPU denotes the elliptically polarized undulator.}
\label{fig:0}
\end{figure}

SHINE is a CW XFEL facility designed to deliver high-brightness electron beams with a typical bunch charge of 100 pC and beam energies up to 8 GeV. It comprises two beamlines—FEL-I and FEL-II—covering photon energy ranges of 3–15 keV and 0.4–3 keV, respectively. Together, these beamlines enable coherent X-ray generation across a broad spectrum from 0.4 keV to 15 keV at repetition rates approaching 1 MHz \cite{Zhao:FEL2017,Liu2023}. In addition to full-energy operation at 8 GeV, the FEL-II beamline can operate with electron beam energies in the range of 3–4.5 GeV, which is particularly suitable for high-repetition-rate BEUV and soft X-ray generation \cite{Liu2023,Yan2021,qi2025}. Figure~\ref{fig:0} shows the schematic layout of the BEUV setup, which follows an existing FEL-II soft X-ray beamline. The electron beam can reach a peak current of approximately 800 A and a normalized slice emittance of 0.2~mm·mrad at the linac exit. It is then transported via a beam bypass line to the undulator line, which consists primarily of planar undulators (PMUs) with a magnetic period of 55~mm and individual lengths of 4~m. The main electron beam and undulator parameters are summarized in Table~\ref{tab:table1}. It is worth noting that, in the baseline design of SHINE, an afterburner section composed of four EPUs is adopted to extend the accessible photon energy range of FEL-II and to enable flexible polarization control, as shown in Fig.~\ref{fig:0}. Each EPU has a period length of 55~mm and a magnetic length of 4~m, and adopts the APPLE-III magnet configuration \cite{Yu2023}.

\begin{table}[!hbt]
    \centering
	\caption{\label{tab:table1}Main parameters of the electron beam and the undulator line for the FEL-II soft X-ray and BEUV beamline of SHINE.}
	\begin{tabular}{lcc}
        \toprule
		Parameters				&Value	&Unit\\
		\midrule
		{\textbf{Electron beam}}&		&\\
		Energy					&4.5		&GeV\\
		Slice energy spread 	&0.015	&\%\\
		Normalized emittance  	&0.2	&mm$\cdot$mrad\\
		Bunch charge			&100	&pC\\
		Peak current (Gaussian) &800	&A\\
		Bunch length (FWHM)			&110		&fs\\
		{\textbf{Undulator line}}&		&\\
		PMU period length		&55	&mm\\
		PMU length      	&4	&m\\
        PMU modules       &32	&\\
        Interspace                &1	&m\\
		EPU period length				&55	&mm\\
		EPU length				&4		&m\\
		EPU modules				&4		&\\
		\bottomrule
	\end{tabular}
\end{table}

\begin{figure}[!htbp]
\centering
\subfigure[\label{fig:1a}]{
\includegraphics[width=0.45\textwidth]{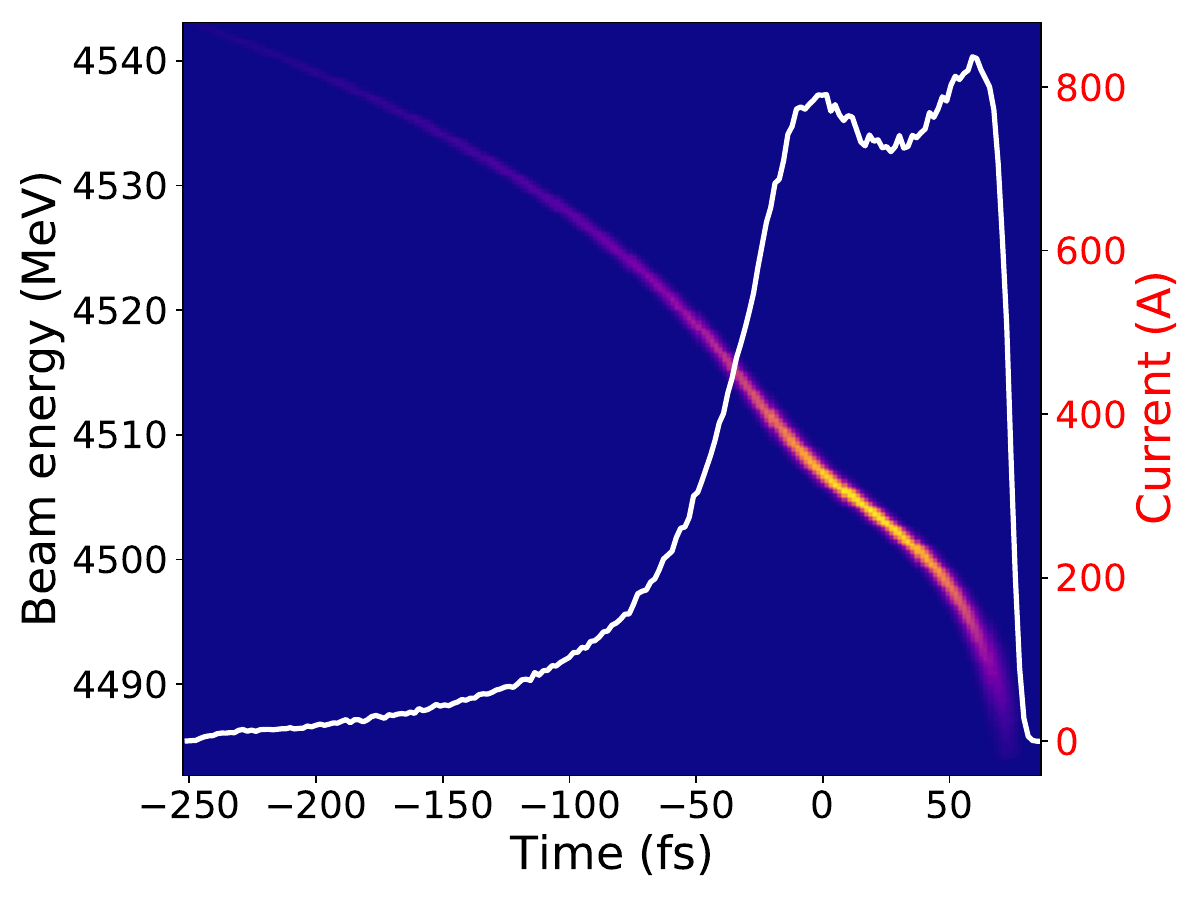}}
\subfigure[\label{fig:1b}]{
\includegraphics[width=0.45\textwidth]{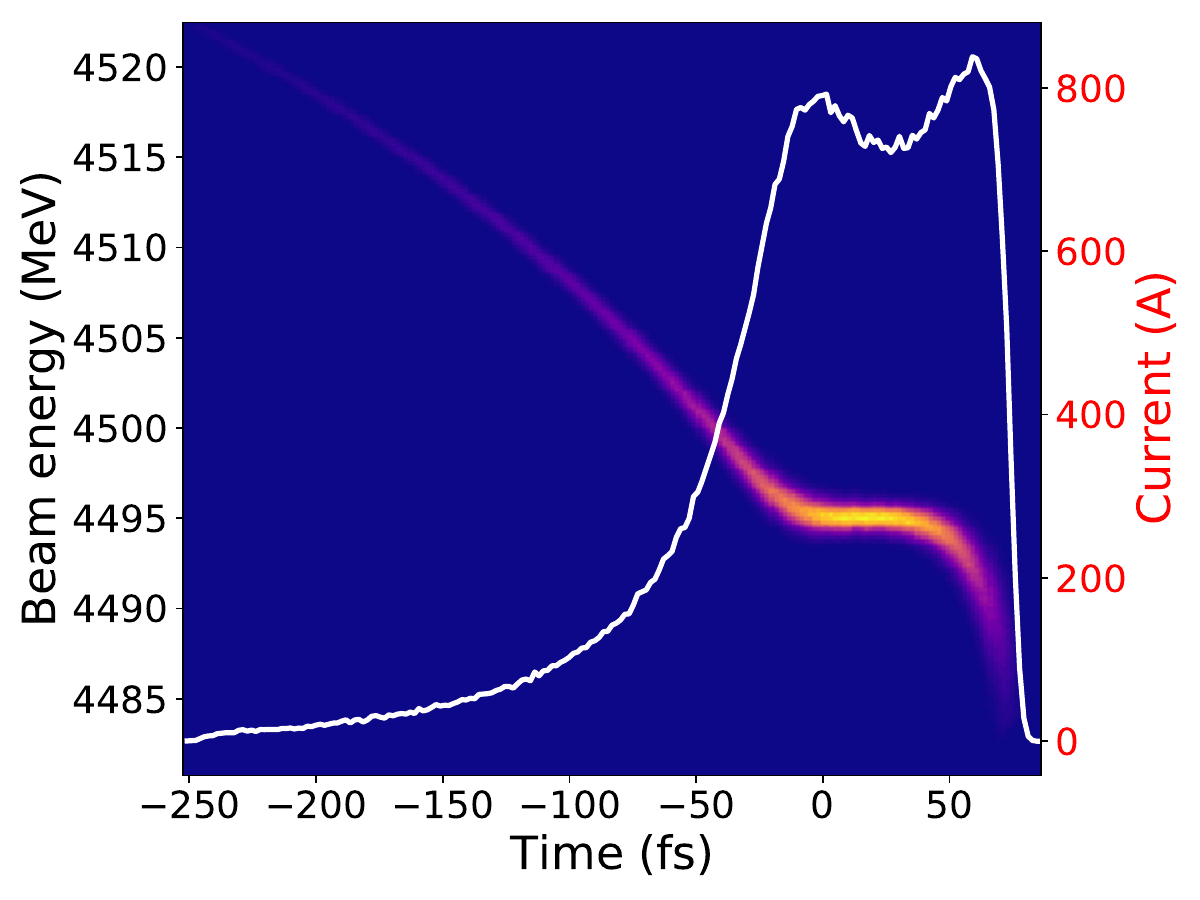}}
\caption{Longitudinal phase space of the electron beam at the undulator entrance (beam head to the right). The color map shows the normalized longitudinal phase-space density, and the white curve represents the current profile. (a) Electron beam with energy chirp. (b) Electron beam after dechirping.}
\label{fig:1}
\end{figure}
To evaluate the BEUV characteristics at SHINE, a realistic electron beam was obtained using the particle tracking codes ASTRA \cite{floettmann2017astra} and ELEGANT \cite{Borland2000}. Figure~\ref{fig:1} shows the longitudinal phase space of the electron beam at the entrance of the undulator. In the chirped case (Fig.~\ref{fig:1a}), a pronounced linear time–energy correlation can be clearly observed, indicating the presence of a significant energy chirp that may lead to longitudinal detuning of the FEL interaction. After passing through the dechirper (Fig.~\ref{fig:1b}), this correlation is largely removed, resulting in a substantially flattened energy distribution along the bunch while preserving the overall current profile. The comparison between the two cases demonstrates that the dechirper effectively suppresses the initial energy chirp, reducing the longitudinal energy variation by more than an order of magnitude. The resulting unchirped beam therefore provides a more favorable initial condition for efficient and stable FEL amplification in the BEUV regime. These beam parameters are chosen to be consistent with the baseline capabilities of SHINE and play a crucial role in determining the achievable performance of the BEUV setup.

\section{FEL simulation results}\label{sec3}
The output BEUV performance is investigated through comprehensive start-to-end numerical simulations. The FEL simulations are performed using the GENESIS code \cite{REICHE1999}. Unless otherwise specified, the reported average FEL power values are calculated based on an assumed electron beam repetition rate of 1 MHz. The baseline BEUV characteristics are first evaluated, followed by a systematic study of the effects of energy chirp and undulator tapering on the FEL process. In addition, the transverse radiation profile and multi-shot stability are analyzed to assess the feasibility of high-average-power output.

\subsection{Baseline BEUV performance}\label{sec:3.1}
The baseline BEUV performance is evaluated for both chirped and unchirped beams to establish a reference for subsequent analyses. These results serve as a benchmark for assessing the impact of beam imperfections and the efficacy of optimization strategies presented later in this section.
\begin{figure}[!htbp]
\centering
\subfigure[\label{fig:2a}]{
\includegraphics[width=0.31\textwidth]{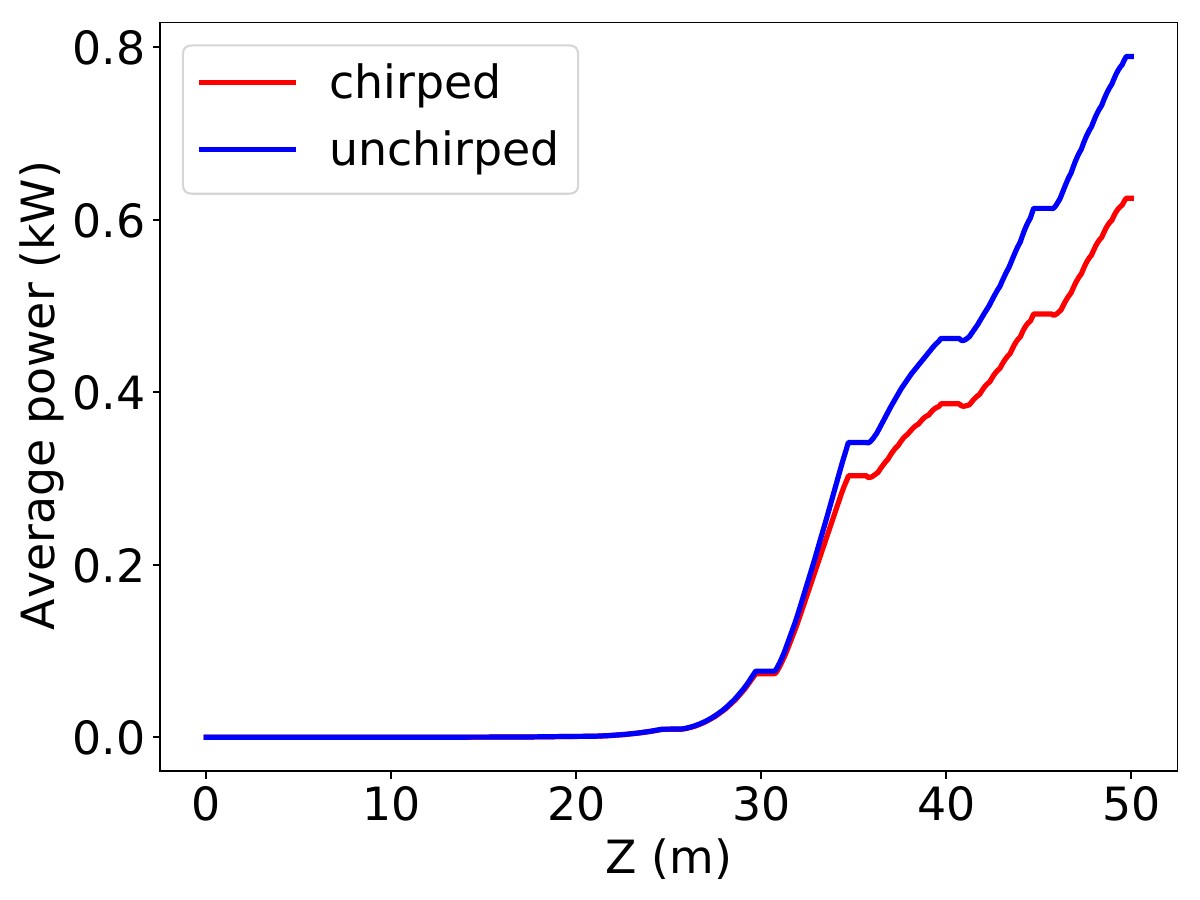}}
\subfigure[\label{fig:2b}]{
\includegraphics[width=0.31\textwidth]{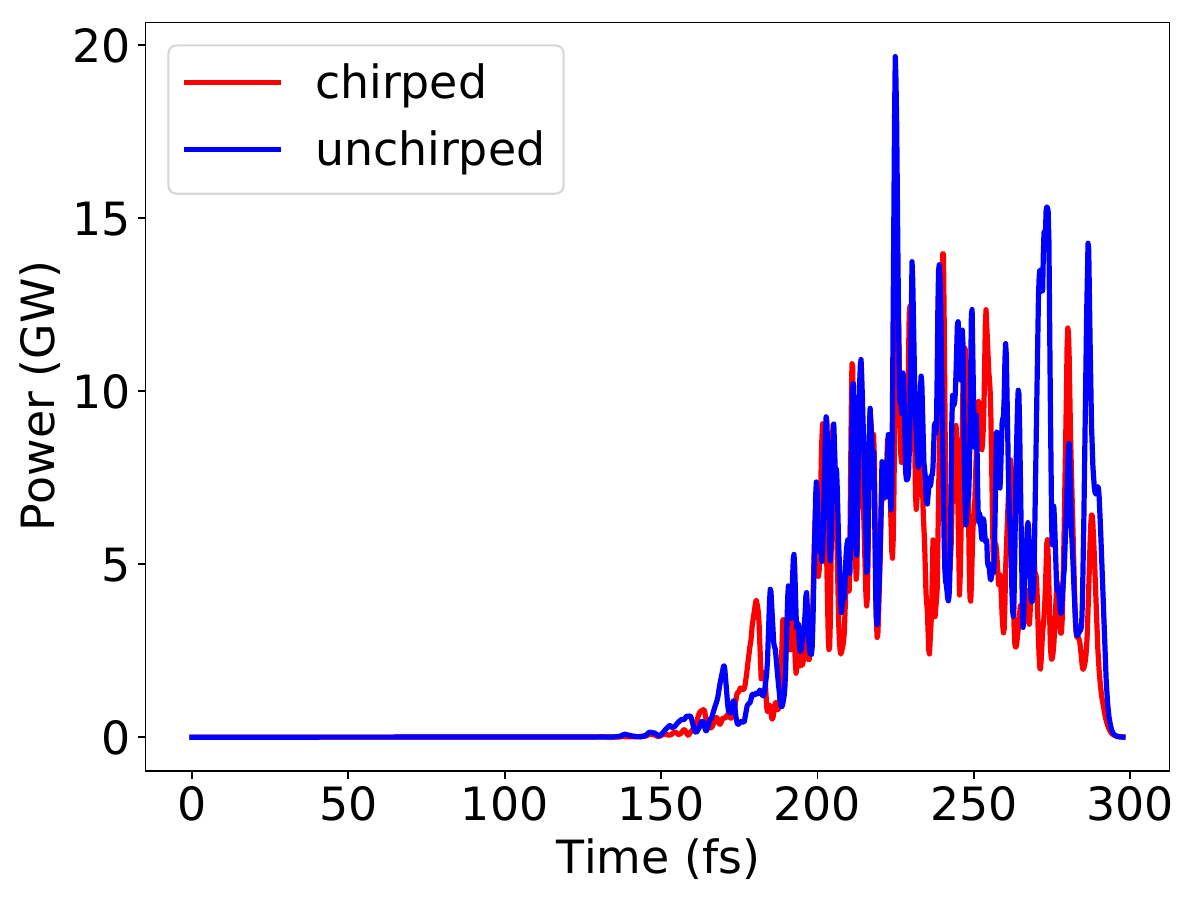}}
\subfigure[\label{fig:2c}]{
\includegraphics[width=0.31\textwidth]{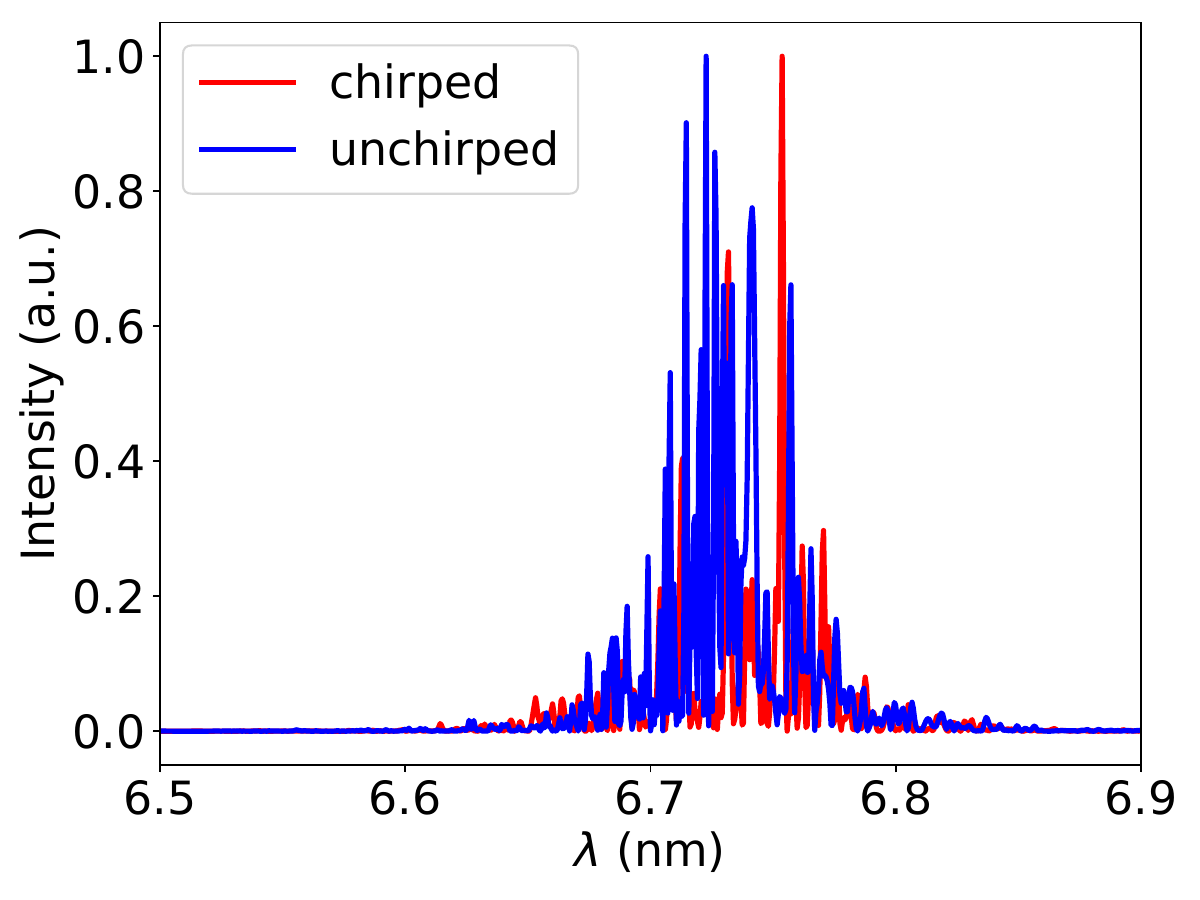}}
\caption{Baseline BEUV performance for chirped (red) and unchirped (blue) electron beams. (a) Evolution of the BEUV power along the undulator. (b) Temporal power profiles at saturation. (c) Corresponding FEL spectra.}
\label{fig:2}
\end{figure}
Figure~\ref{fig:2} presents the BEUV performance in the absence of undulator tapering. The results are obtained using the same undulator configuration, lattice, and initial shot noise for both the chirped and unchirped cases, and are intended as a baseline comparison, where the observed differences primarily reflect the effect of the energy chirp rather than shot noise fluctuations. The saturated BEUV pulse length and average power are comparable in the two cases. For the chirped beam, the BEUV reaches an average power of 0.62~kW with a pulse duration of 80.2 fs, while the unchirped case yields an average power of 0.79~kW and a slightly shorter pulse duration of 76.2~fs. The corresponding spectral properties show that the rms bandwidth is 0.59\% for the chirped beam and 0.49\% for the unchirped beam. Although the presence of a linear energy chirp leads to a modest spectral broadening, the impact on the overall spectral quality remains limited under baseline BEUV operation. This indicates that, at the baseline operating point, the intrinsic FEL bandwidth dominates over chirp-induced broadening for the level of energy chirp considered here. In particular, the resulting bandwidth in both cases is well below 1\%, which satisfies the bandwidth requirements of BEUV multilayer mirrors \cite{nano11112782,endo2014extendibility}.

\subsection{Effects of energy chirp and undulator tapering}
\begin{figure}[!htbp]
\centering
\includegraphics[width=0.45\textwidth]{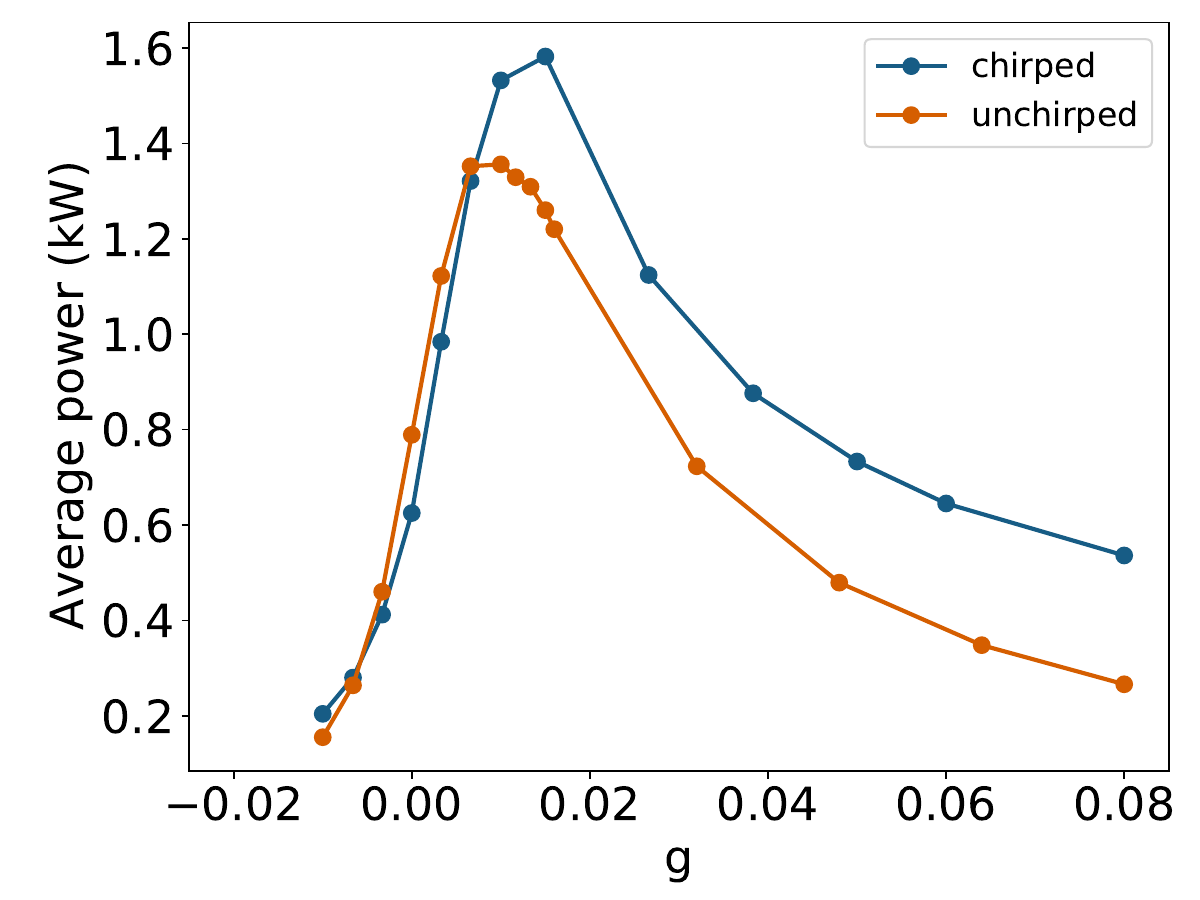}
\caption{Output average power of the BEUV FEL as a function of the linear undulator taper strength $g$ for chirped and unchirped beams.}
\label{fig:3}
\end{figure}
The beam exhibits a relative linear energy chirp of approximately $2.2\times10^{-3}$ over the high-current region, which is comparable to the FEL parameter $\rho = 2.1\times10^{-3}$ calculated using the Ming-Xie formulas \cite{XIE200059,Huang2007}. Consequently, the energy chirp has no significant impact on the spectral bandwidth in this case. Moreover, when combined with a linearly tapered undulator, this chirped beam can not only compensate for slippage-induced efficiency degradation but also enhance the output power beyond that of an unchirped beam \cite{Saldin2006, SCHNE2022}. Typically, in a SASE FEL, the following condition is satisfied:
\begin{equation}
    \frac{dK}{dz} = -\frac{(1 + K_0^2/2)^2}{K_0} \frac{1}{\gamma_0^3} \frac{d\gamma}{c\,dt},
\end{equation}
where $\gamma$ denotes the relativistic Lorentz factor of the electron beam, $\gamma_0$ is the Lorentz factor corresponding to the FEL resonance condition, $K$ is the undulator field parameter, and $K_0$ is its value at resonance. When $\rho \to 0$, the linear undulator tapering can nearly perfectly compensate for the gain degradation induced by the linear energy chirp in the SASE FEL process \cite{SCHNE2022}. In this case, the induced resonant wavelength shift becomes comparable to or smaller than the FEL bandwidth, allowing a much larger fraction of the electron bunch to remain resonant and contribute coherently to the FEL gain.

To further optimize the FEL performance, we investigate the effect of linear undulator tapering in PMUs for both chirped and unchirped beams. The taper strength $g$ can be described as:
\begin{equation}
    K(z) = K_0 \left(1 - g \frac{z}{L}\right).
\end{equation}
where $L$ is the total undulator length. In this study, the taper is introduced starting from $z$ = 30 m, corresponding to the pre-saturation regime and aligned with an undulator module boundary. Figure~\ref{fig:3} shows the FEL average power at $z$ = 50 m as a function of the linear undulator tapering. In the absence of an energy chirp, tapering leads to a moderate enhancement of the average power, with an optimal output of approximately 1.3~kW at $g = 0.005$. In contrast, for the chirped beam, a properly designed taper significantly improves the energy extraction efficiency, resulting in a higher BEUV output power of about 1.6 kW at $g = 0.015$.

The enhanced FEL performance observed with the chirped beam arises from the interplay between the longitudinal energy chirp and the undulator tapering. The linear energy chirp induces a gradual shift in the local resonance condition along the electron bunch. When matched with an appropriately undulator-tapering, this shift can be partially compensated, enabling a larger fraction of the bunch to remain in resonance over an extended interaction length. Consequently, energy extraction becomes more efficient than in the unchirped case. These results demonstrate that, rather than being purely detrimental, a controlled energy chirp—when combined with optimized undulator tapering—can be leveraged to enhance FEL output power.



\subsection{Transverse profile}
\begin{figure}[!htbp]
\centering
\includegraphics[width=0.55\textwidth]{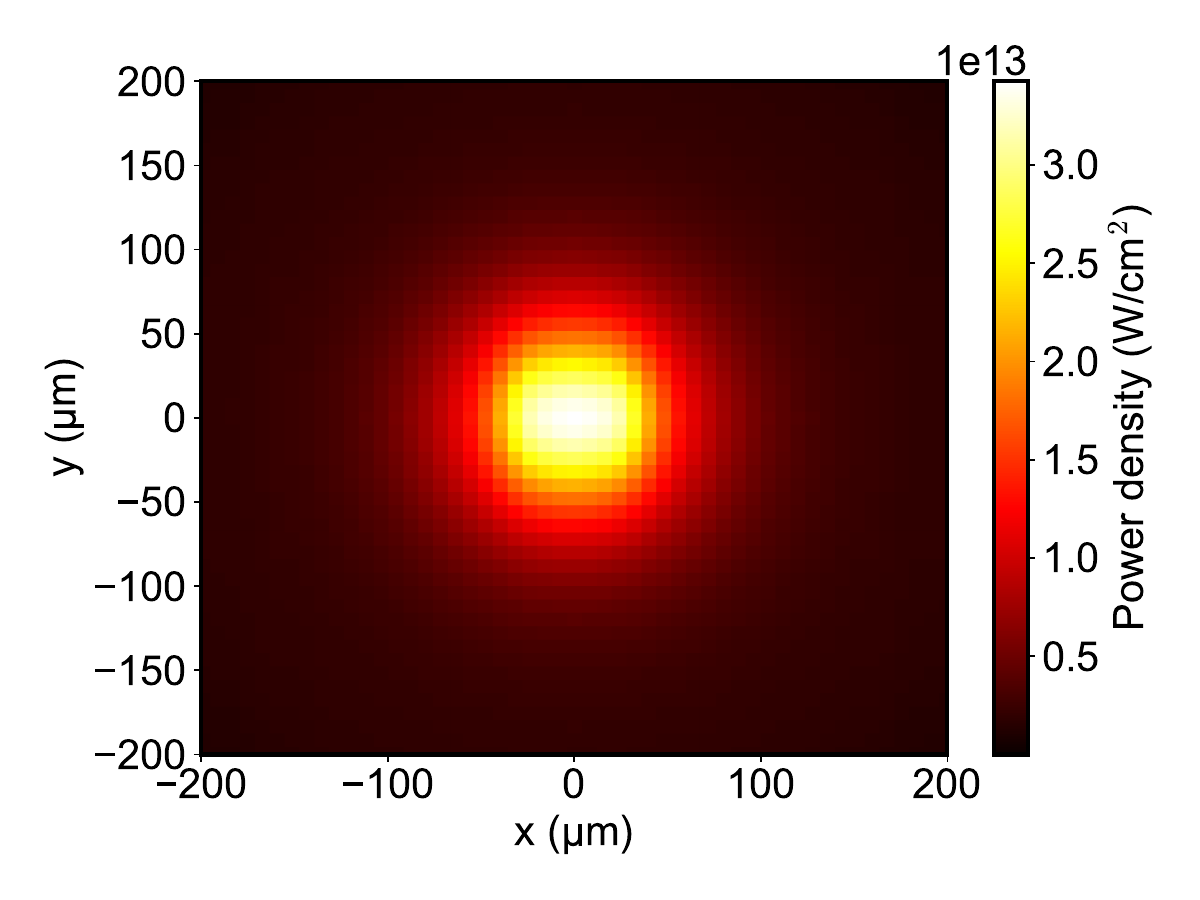}
\caption{Transverse power density distribution of the BEUV radiation at saturation.}
\label{fig:4}
\end{figure}
Beyond the total output power, the transverse power density distribution of the BEUV radiation is a critical factor for lithographic applications. It determines the peak energy loading on optical components, the transverse mode quality, and the effective utilization of the BEUV-FEL radiation. Figure~\ref{fig:4} shows the transverse power density distribution of the BEUV FEL radiation at saturation, obtained for the unchirped beam under the optimized undulator tapering conditions. The radiation exhibits a nearly Gaussian transverse profile with a well-confined central core, indicating good transverse coherence and a high-quality fundamental transverse mode. The full width at half maximum (FWHM) of the transverse spot is approximately 144~$\mu$m. The peak power density reaches about $3.3\times10^{13}\,\mathrm{W/cm^{2}}$ and the energy density (dose per pulse) is around $2.5\times10^{3}\,\mathrm{mJ/cm^{2}}$.

The Gaussian-like transverse profile also implies that the BEUV radiation is fully transversely coherent, which is a natural property of SASE FELs. For BEUV lithography, the availability of such a high-intensity source is particularly attractive, since the high photon flux can effectively suppress stochastic noise arising from limited photon dose, thereby improving the optical contrast in the photoresist. At the same time, the strong transverse coherence of BEUV radiation poses challenges for illumination uniformity in lithographic systems. This issue can be addressed by advanced illumination schemes based on transverse beam-splitting techniques \cite{Tomas2024}. Such schemes, which have already been demonstrated for EUV FELs, decompose a single FEL pulse into multiple beamlets. Owing to millimeter-scale path-length differences between different illumination channels, the overall exposure becomes effectively incoherent, even though each individual beamlet remains transversely coherent. This provides a practical route for managing the coherence properties of BEUV radiation while fully benefiting from its high peak and average power.

\subsection{Multi-shot stability analysis}
For high-repetition-rate operation, shot-to-shot stability is a key performance metric. To evaluate the robustness of the BEUV setup at SHINE, we perform multi-shot FEL simulations and analyze the fluctuations in pulse energy and spectral properties. The results provide insight into the stability of the FEL output under realistic operating conditions.

\begin{figure}[!htbp]
\centering
\subfigure[\label{fig:5a}]{
\includegraphics[width=0.45\textwidth]{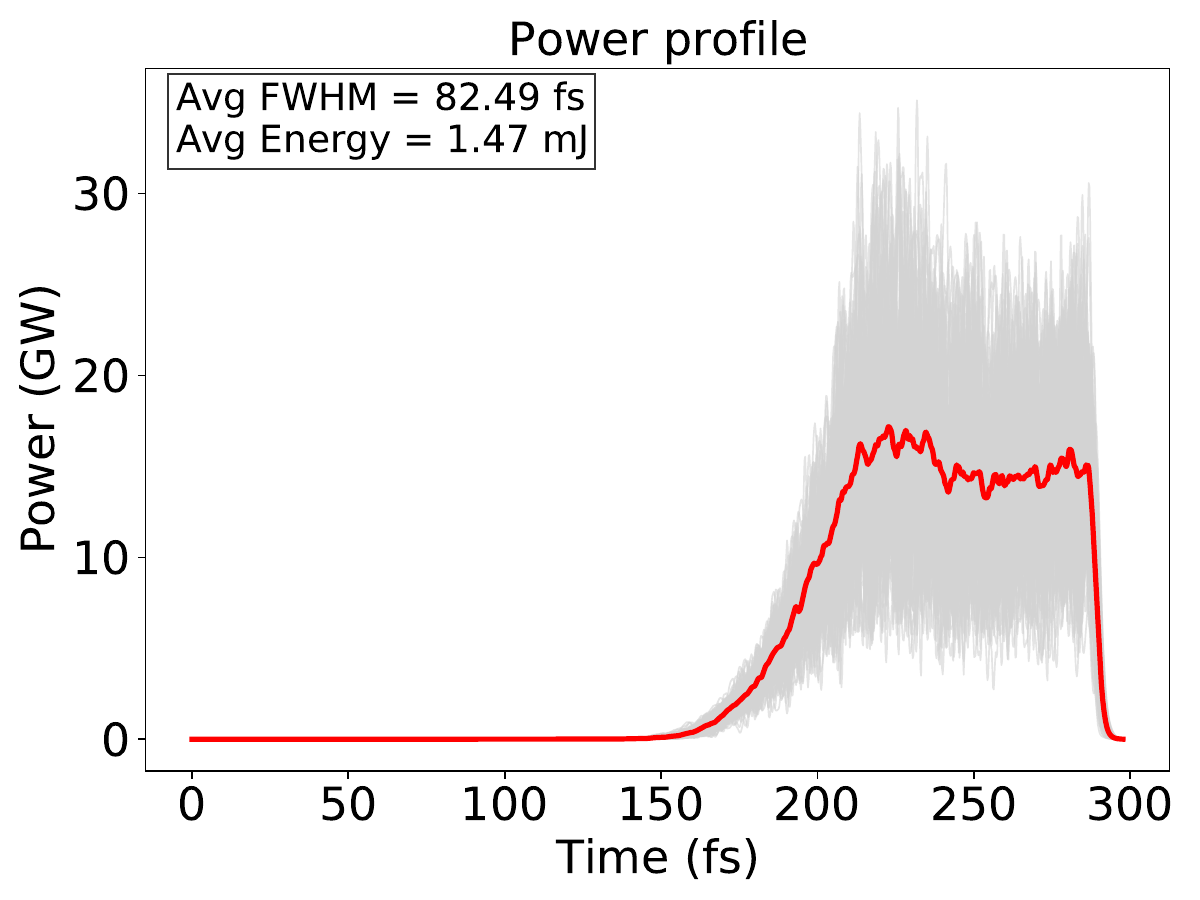}}
\subfigure[\label{fig:5b}]{
\includegraphics[width=0.42\textwidth]{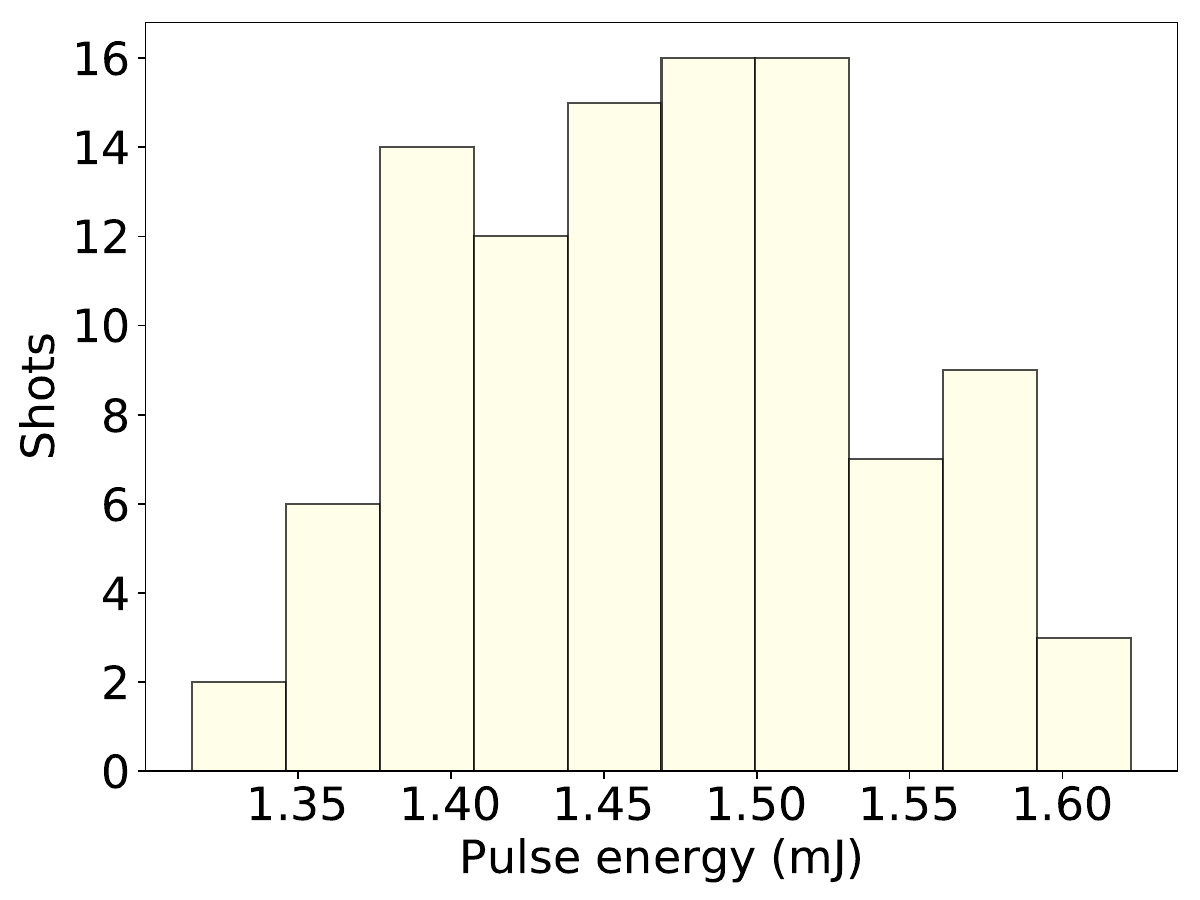}}
\subfigure[\label{fig:5c}]{
\includegraphics[width=0.45\textwidth]{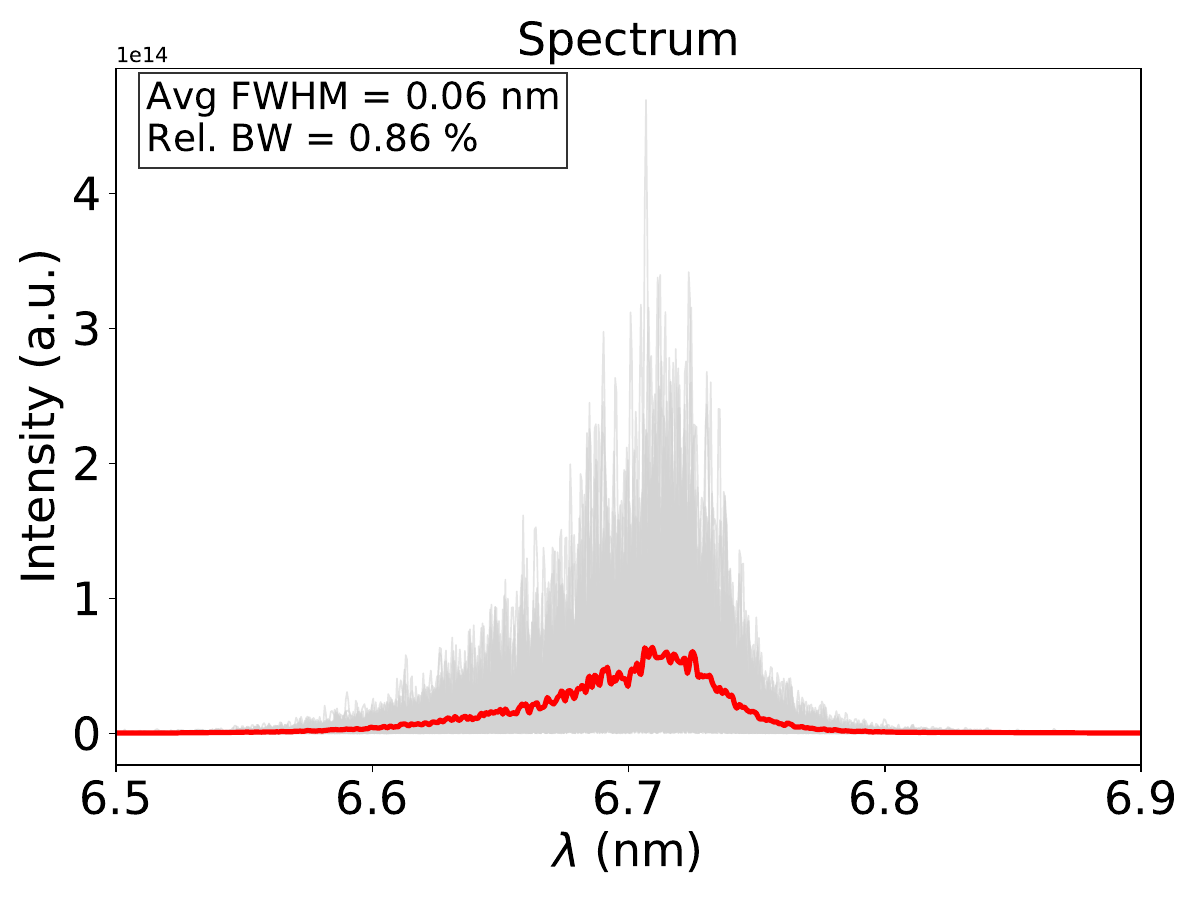}}
\subfigure[\label{fig:5d}]{
\includegraphics[width=0.42\textwidth]{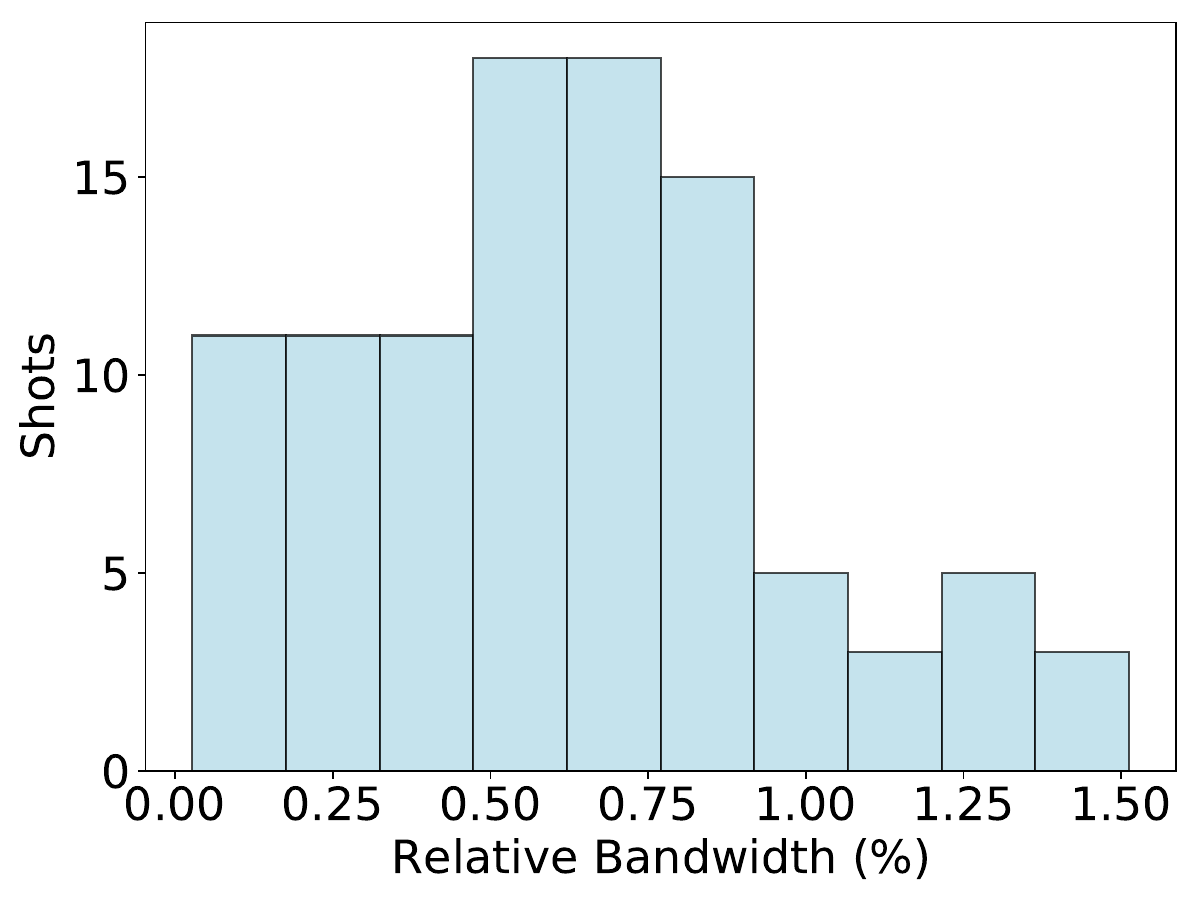}}
\caption{Shot-to-shot performance of the BEUV radiation for the chirped beam under optimized undulator tapering conditions, obtained from 100-shot simulations. 
(a) Temporal power profiles, where the red curve represents the averaged profile and the gray curves show individual shots. 
(b) Histogram of the pulse energy distribution. 
(c) Radiation spectra, with the red curve showing the averaged spectrum and the gray curves indicating single-shot spectra. 
(d) Histogram of the relative spectral bandwidth.}
\label{fig:5}
\end{figure}

\begin{figure}[!htbp]
\centering
\subfigure[\label{fig:6a}]{
\includegraphics[width=0.45\textwidth]{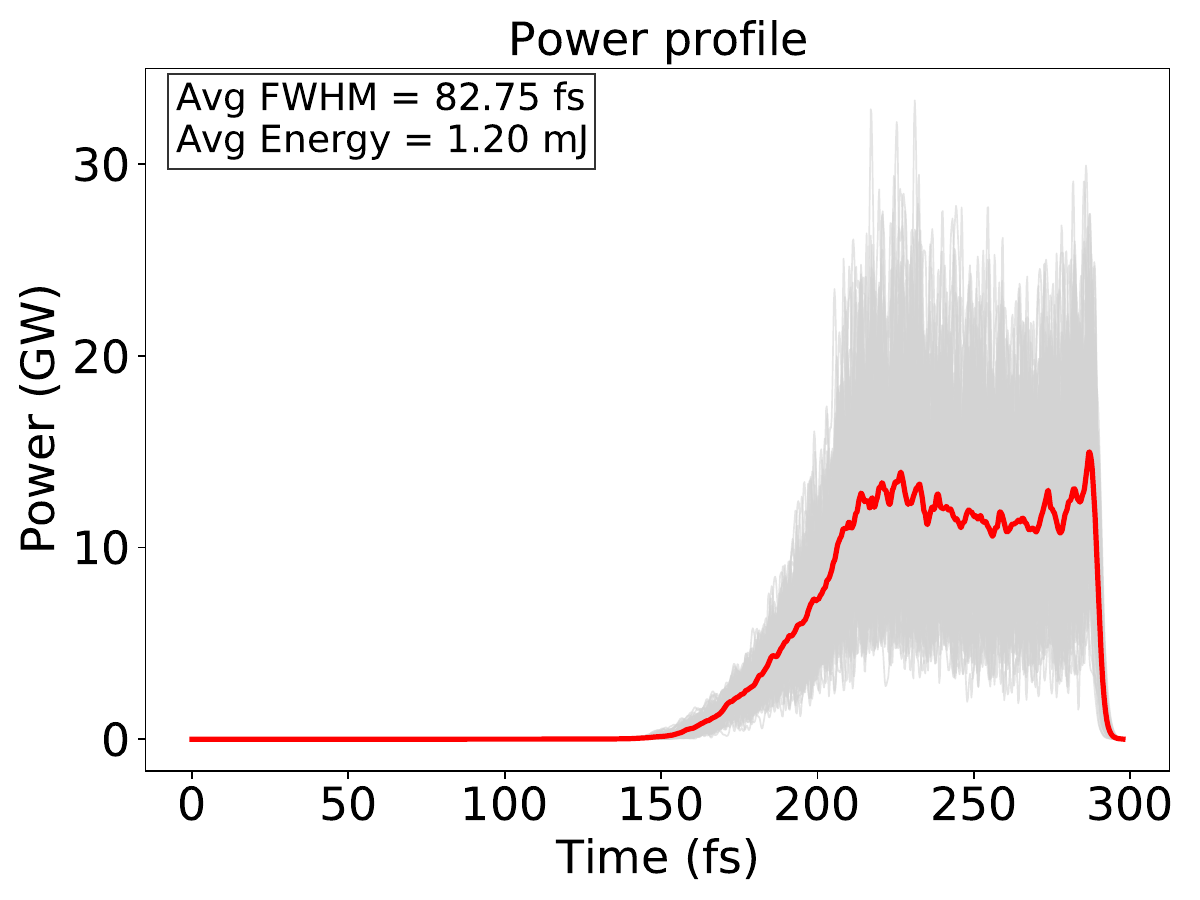}}
\subfigure[\label{fig:6b}]{
\includegraphics[width=0.42\textwidth]{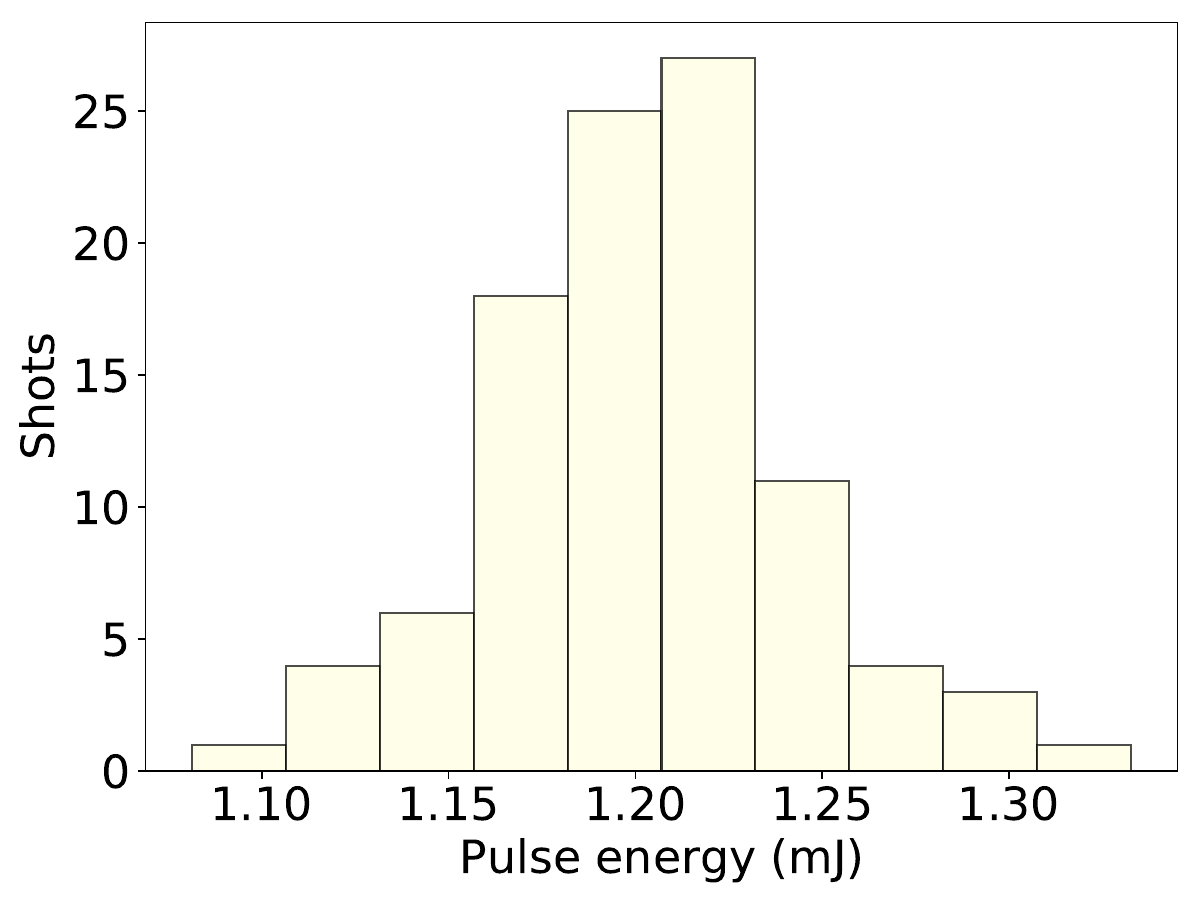}}
\subfigure[\label{fig:6c}]{
\includegraphics[width=0.45\textwidth]{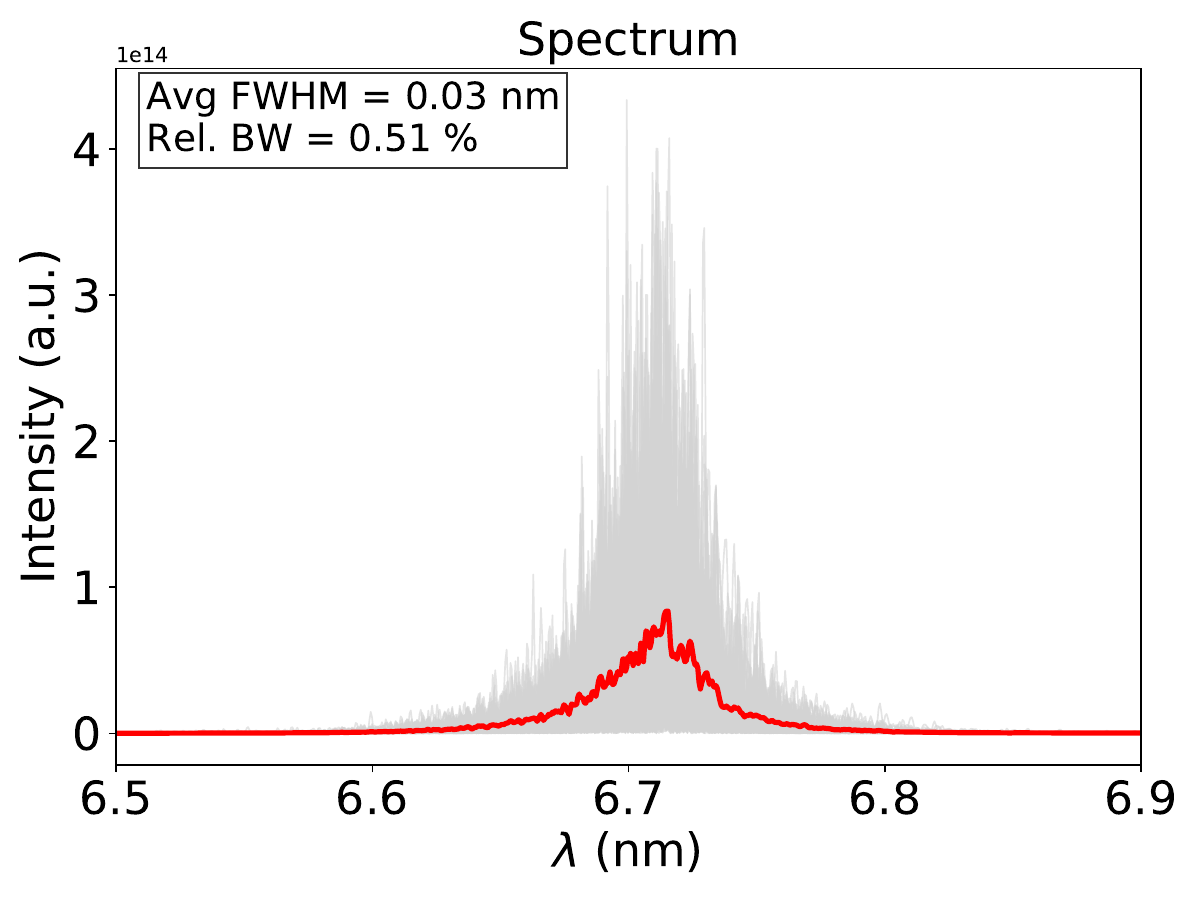}}
\subfigure[\label{fig:6d}]{
\includegraphics[width=0.42\textwidth]{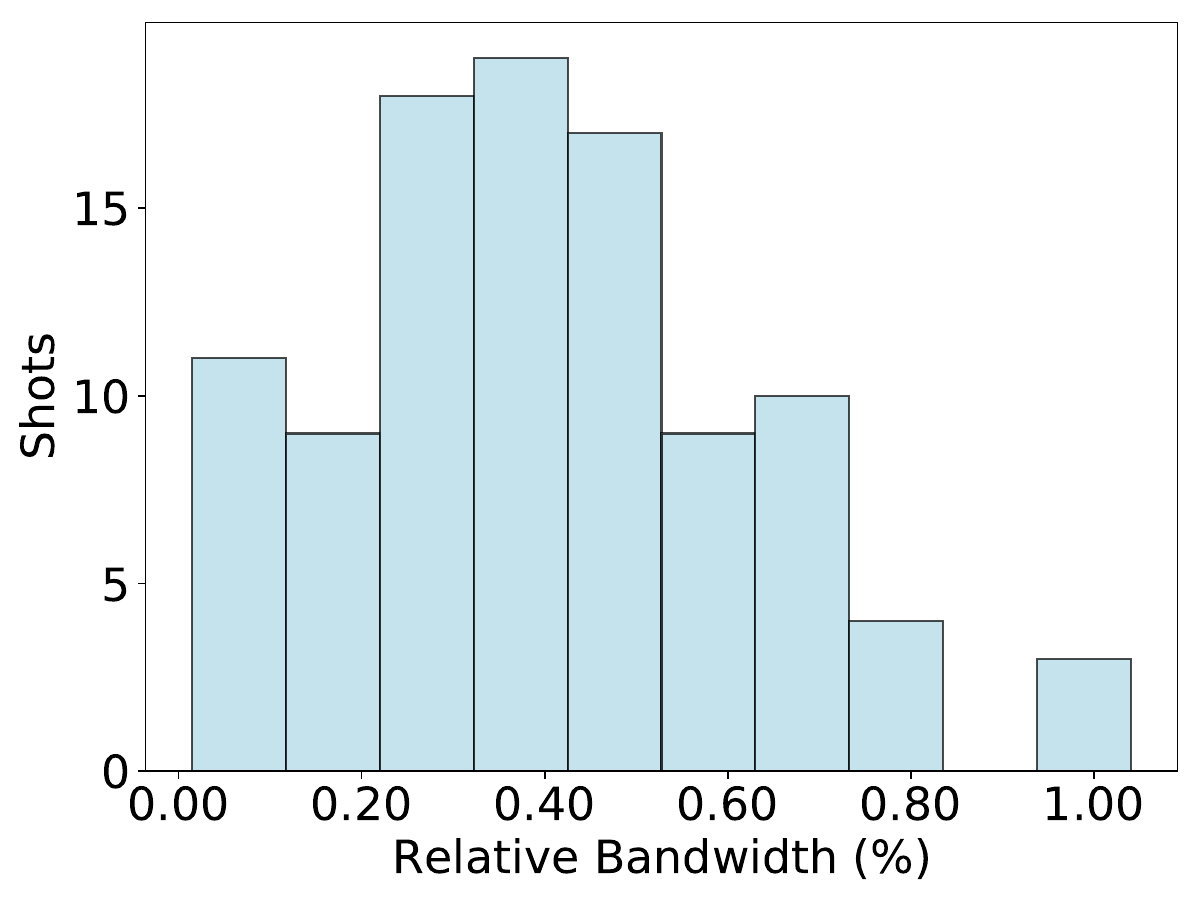}}
\caption{Shot-to-shot performance of the BEUV radiation for the unchirped beam under optimized undulator tapering conditions, obtained from 100-shot simulations. 
(a) Temporal power profiles, where the red curve represents the averaged profile and the gray curves show individual shots. 
(b) Histogram of the pulse energy distribution. 
(c) Radiation spectra, with the red curve showing the averaged spectrum and the gray curves indicating single-shot spectra. 
(d) Histogram of the relative spectral bandwidth.}
\label{fig:6}
\end{figure}

The 100-shot statistical properties are summarized in Fig.~\ref{fig:5} and \ref{fig:6}, where the temporal profiles, pulse energy distributions, radiation spectra, and relative bandwidth distributions are presented for both the chirped and unchirped cases under optimized tapering conditions. For the chirped beam, the 100-shot statistics show an average pulse energy of 1.47 mJ with a standard deviation of 68.1 $\mu$J, corresponding to a relative fluctuation of 4.6\%. The average FWHM pulse duration is 82.5 fs with a standard deviation of 5.1 fs, giving a relative fluctuation of 6.2\%. The rms spectral bandwidth has an average value of 0.62\%, while its relative fluctuation reaches 56.4\%. The spectrum obtained by averaging over 100 shots exhibits a FWHM relative bandwidth of 0.86\%, which provides a more representative characterization of the SASE spectral bandwidth.

For the unchirped beam, the average pulse energy is 1.20 mJ with a standard deviation of 41.4 $\mu$J, corresponding to a smaller relative fluctuation of 3.4\%. The average FWHM pulse duration is 82.8 fs with a standard deviation of 5.5 fs, yielding a relative fluctuation of 6.7\%. The average rms spectral bandwidth is reduced to 0.40\%, with a relative fluctuation of 55.1\%. The corresponding multi-shot averaged spectrum has a FWHM relative bandwidth of 0.51\%, again providing a direct measure of the effective SASE spectral bandwidth.

These results indicate that, under optimized tapering conditions, both operation modes exhibit good temporal stability, with comparable pulse durations and similar relative fluctuations at the level of about 6–7\%. The shot-to-shot energy stability is also well maintained in both cases, with relative jitters below 5\%. The unchirped beam provides slightly better pulse energy stability and a narrower average bandwidth, whereas the chirped beam delivers a significantly higher pulse energy, increased by about 22\% compared with the unchirped case. This directly translates into a higher achievable average BEUV output power at a given repetition rate.

The spectral bandwidth shows relatively large shot-to-shot fluctuations for both cases, which is characteristic of SASE FEL operation. Nevertheless, the average rms bandwidths remain well below 1\% in both modes, satisfying the requirements of BEUV multilayer optics and lithographic applications. The larger mean bandwidth in the chirped case reflects the additional longitudinal energy variation of the electron beam, while its higher pulse energy confirms that the optimized undulator tapering effectively exploits the chirp to enhance the energy extraction efficiency.

Overall, the comparison demonstrates a clear trade-off between output power and spectral purity. The chirped-beam operation, when combined with optimized tapering, favors higher pulse energy and thus higher average BEUV power, making it particularly attractive for applications demanding maximum photon flux, such as BEUV lithography. In contrast, the unchirped-beam operation provides slightly improved spectral quality and marginally better energy stability, which may be advantageous for applications requiring higher spectral precision. Both modes therefore represent viable and complementary operation regimes for a high-average-power BEUV setup at SHINE.

\section{BEUV polarization control}\label{sec4}
\begin{figure}[htbp]
\centering
\subfigure[\label{fig:7a}]{
\includegraphics[width=0.31\textwidth]{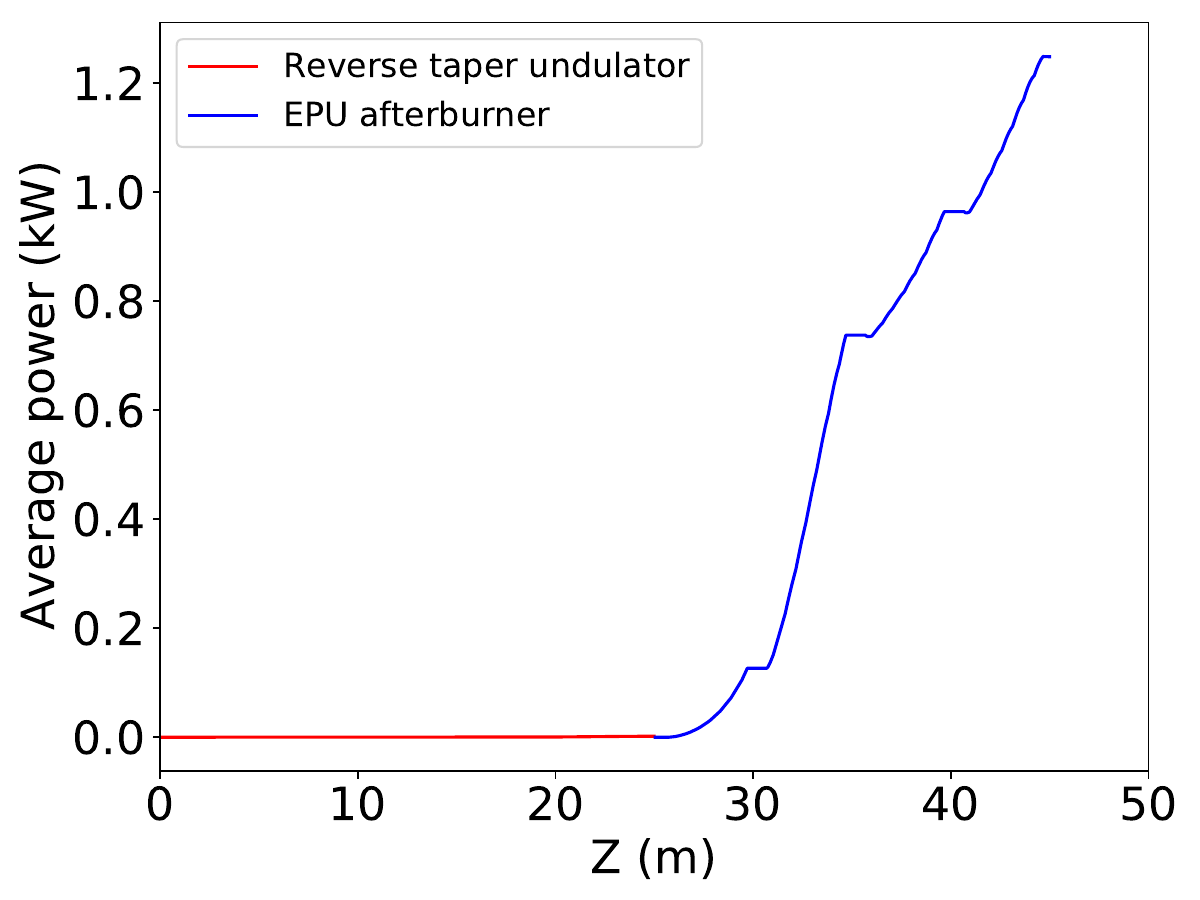}}
\subfigure[\label{fig:7b}]{
\includegraphics[width=0.31\textwidth]{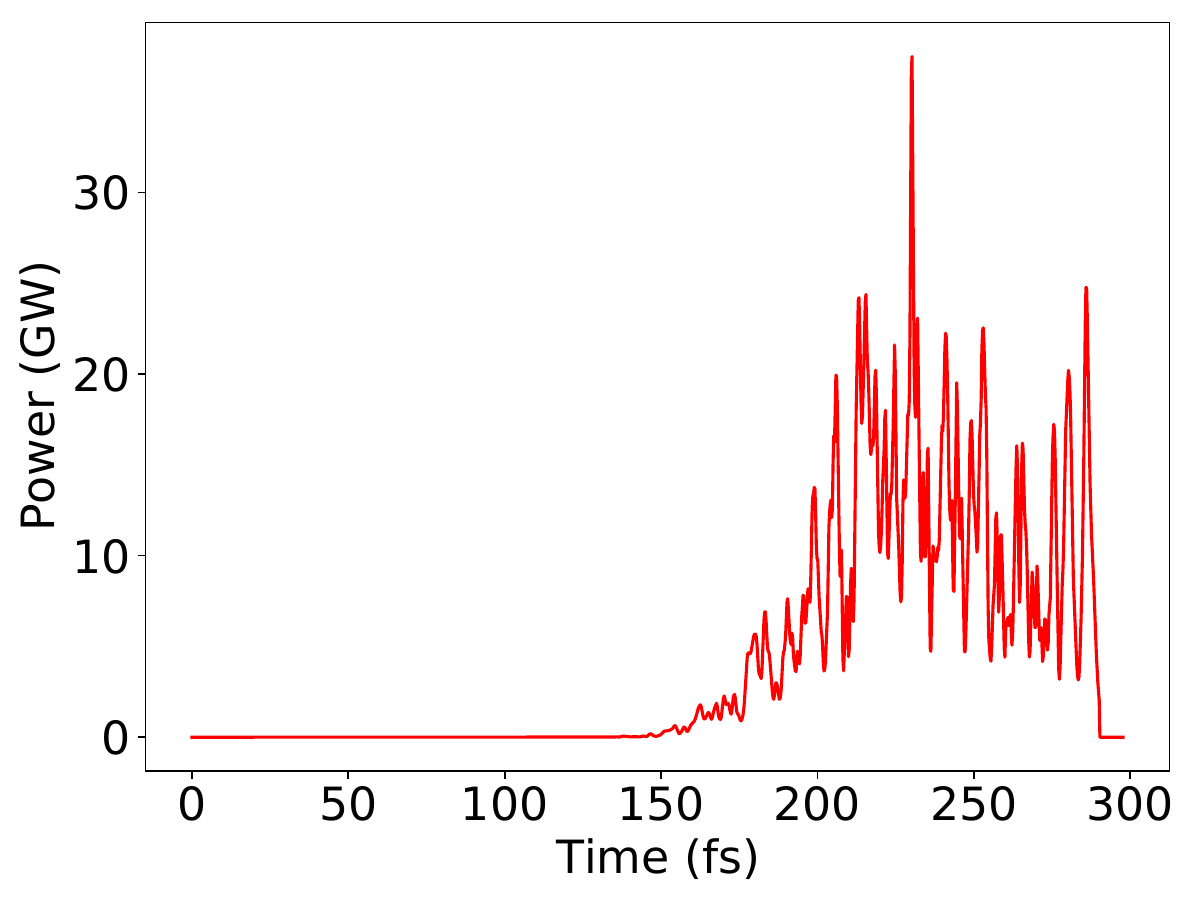}}
\subfigure[\label{fig:7c}]{
\includegraphics[width=0.31\textwidth]{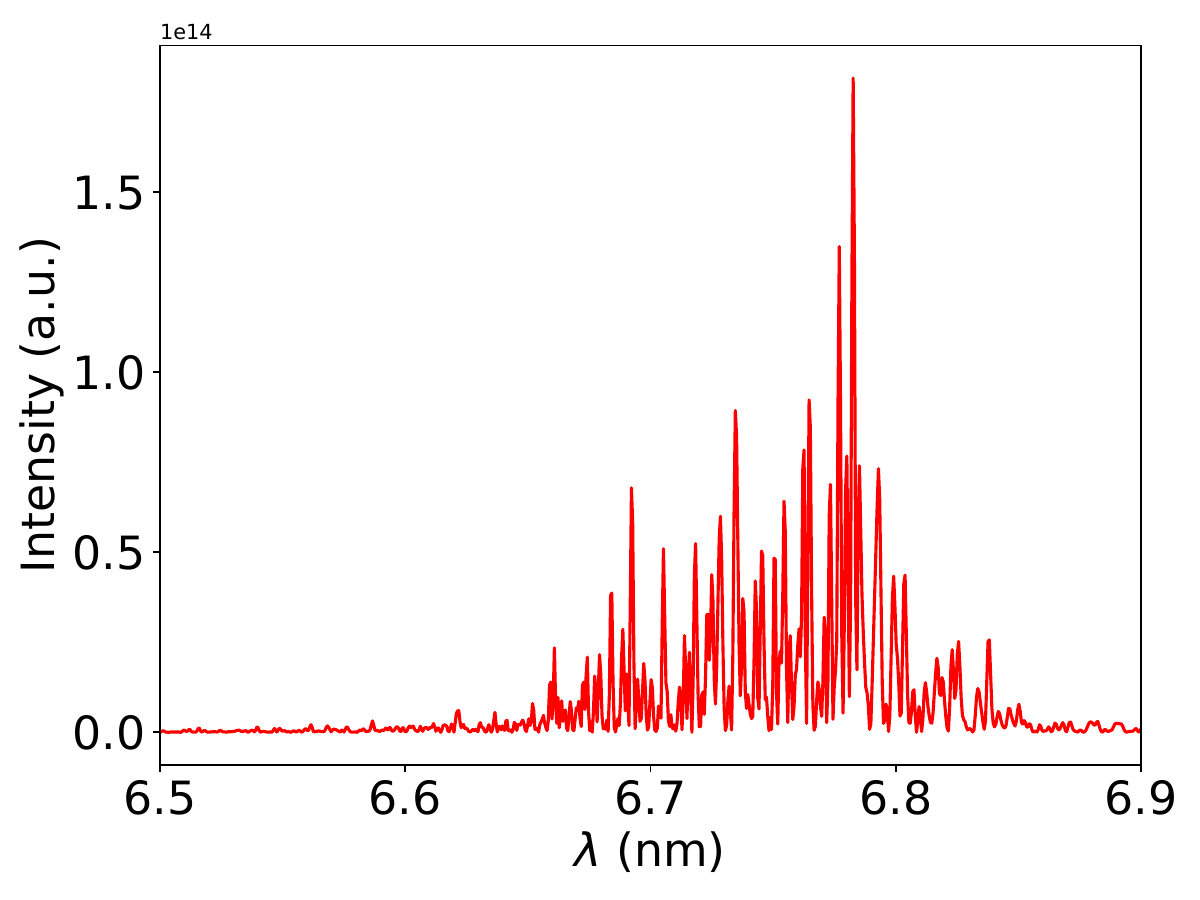}}
\caption{BEUV performance of the EPU afterburner for generating circularly polarized BEUV radiation. (a) BEUV average power evolution along the undulator line. (b) Power profile of the circularly polarized BEUV radiation at the exit of the EPUs. (c) Corresponding radiation spectrum.}
\label{fig:7}
\end{figure}
The polarization properties of the BEUV light source are of particular importance for lithography applications \cite{Uschakow_2013,levinson2022high}. For high-numerical-aperture (high-NA) lithography, $s$-polarized (TE-polarized) radiation provides superior image contrast and intensity compared with $p$-polarized (TM-polarized) radiation. In contrast, for low-NA lithography, the performance of $s$- and $p$-polarized light is nearly identical \cite{Bruce2004}. Since hyper-NA lithography is associated with substantially higher cost and technical complexity, we consider two complementary operation modes in the BEUV setup at SHINE: a high-average-power linearly polarized radiation based on the planar undulators for high-NA lithography, and a circularly polarized radiation based on the EPU afterburner for low-NA lithography with higher total average power.

Planar undulators are typically employed to generate linearly polarized radiation in SASE FELs. Other polarization states, such as elliptical and circular polarization, can be directly obtained by using EPUs, as demonstrated at facilities such as FERMI \cite{allaria2012highly} and SwissFEL \cite{milne2017}. In addition, arbitrary polarization states can be generated by superposing two orthogonal linearly polarized radiation fields produced by crossed planar undulators combined with phase shifters \cite{ferrari2019free}. However, due to the lack of longitudinal coherence in SASE FELs, this crossed-undulator approach generally results in a limited degree of polarization, whereas much higher polarization purity can be achieved in externally seeded FELs \cite{Emaa2014,Deng2014de}.

SHINE adopts a flexible EPU afterburner configuration to enable polarization control and rapid switching of the FEL radiation. The EPU afterburner can be operated in two distinct modes. In the first mode, fast phase shifters installed between adjacent undulator segments enable the formation of crossed planar undulators, allowing rapid switching between different polarization states at high repetition rates \cite{KIM1984425,KIM2000329}. In the second mode, a reverse-taper scheme is applied to the upstream planar undulators \cite{lutman2016polarization}. In this configuration, the undulator parameter is gradually increased along the undulator line, which suppresses the growth of FEL radiation power while still allowing the electron beam to develop strong microbunching. As a result, the electron beam exits the planar undulator with significant bunching but relatively low radiation power. The pre-bunched beam then enters the downstream EPU afterburner, where efficient radiation emission is restored, and the polarization state is determined by the EPU configuration, enabling the generation of high-power circularly polarized FEL radiation. Compared with conventional approaches relying solely on planar or helical undulators, this scheme offers a simplified method to achieve flexible polarization control while maintaining high output power. In particular, it avoids strong competition between polarization modes in the main radiator and allows the polarization properties to be defined primarily in the afterburner section. It should be noted that the effectiveness of this scheme depends on the preservation of microbunching through the reverse-taper section and may introduce additional sensitivity to beam quality and tapering optimization.

As illustrated in Fig.~\ref{fig:7}, an unchirped electron beam passes through a 25~m long planar undulator operated in a reverse-taper mode with a taper strength of $g=-0.007$, which effectively suppresses the generation of linearly polarized radiation while preserving the electron microbunching. Under this condition, the residual FEL pulse energy from the planar undulators is only about 1.9~$\mu$J. The microbunched beam then enters the EPU afterburner, where the pulse energy is amplified to approximately 1.25~mJ after four EPU segments. At a repetition rate of 1 MHz, this corresponds to an average power of about 1.25~kW, which is nearly twice the baseline BEUV FEL average power of approximately 0.8~kW, as discussed in Sec.~\ref{sec:3.1}. The resulting FEL pulse duration is approximately 80.6~fs, with an rms spectral bandwidth of 0.97\%.

For a high degree of circular polarization, the polarization degree can be estimated by \cite{gao2020polarization}
\begin{equation}
D_{\mathrm{cir}} \approx 1 - \frac{P_{\mathrm{lin}}}{2P_{\mathrm{cir}}},
\end{equation}
where $P_{\mathrm{lin}}$ and $P_{\mathrm{cir}}$ denote the intensities of the linearly and circularly polarized radiation components, respectively. Based on the simulated radiation powers, the circular polarization degree is evaluated to be as high as 99.9\%.

With an electron beam repetition rate of 1~MHz, the corresponding average power of the circularly polarized BEUV radiation can reach approximately 1.3~kW. This demonstrates that the BEUV setup provides a promising route toward a high-average-power, highly circularly polarized BEUV radiation, which is well suited for lithography and other applications requiring flexible polarization control.

\section{Conclusion}\label{sec5}
A high-power BEUV setup at SHINE has been systematically studied. By exploiting the high beam energy and repetition rate of the SRF linac and applying undulator tapering to enhance the extraction efficiency, kilowatt-level average BEUV radiation at MHz repetition rates is shown to be achievable. The effects of electron beam energy chirp and taper optimization on the FEL performance are quantitatively analyzed. In addition, a practical polarization control scheme combining reverse-taper PMUs and an EPU afterburner is demonstrated, enabling flexible generation of polarized BEUV radiation. These results indicate that SHINE provides a realistic and promising platform for a high-average-power BEUV light source, with strong potential for next-generation lithography and related applications.

\backmatter

\section*{Acknowledgements}
The authors would like to acknowledge S. Z., T. L., D. G., and G. W. of SARI, J. Y. of DESY for their valuable discussions.

\subsection*{Author Contributions}
H. D. conceived the idea and supervised the project. H. Y. performed the FEL simulations and drafted the manuscript. Z. G., B. Y., and W. C. contributed to the FEL and beam dynamics simulations, respectively, and N. H. provided the parameter optimization strategy. All authors discussed the results and contributed to the manuscript revision.

\subsection*{Funding}
This work was supported by the National Key Research and Development Program of China (2024YFA1612101), the National Natural Science Foundation of China (12125508, 12541503), the Shanghai Pilot Program for Basic Research – Chinese Academy of Sciences, Shanghai Branch (JCYJ-SHFY-2021-010), the Innovation Program of Shanghai Advanced Research Institute, CAS (2025CP006), and the China Postdoctoral Science Foundation (2025M770914).

\subsection*{Data availability}
The data that support the findings of this study are not publicly available at this time but may be obtained from the authors upon reasonable request.


\bibliography{sn-bibliography}

\end{document}